\documentclass[showpacs,prl,twocolumn,aps]{revtex4}

\pdfoutput=1

\usepackage[colorlinks,urlcolor=blue,citecolor=blue]{hyperref}

\usepackage{amsmath}
\usepackage{amssymb}
\usepackage{latexsym}
\usepackage{amsfonts}
\usepackage{epsfig}
\usepackage{psfrag}
\usepackage{graphicx}

\def\hhref#1{\href{http://arxiv.org/abs/#1}{arXiv:#1}} 

\usepackage{color}

\newcommand{\bea}{\begin{eqnarray}}
\newcommand{\ea}{\end{eqnarray}}

\begin{document}  

\title{Convergence from Divergence}

\author{Ovidiu Costin$^{1}$ and Gerald~V.~Dunne$^{2}$}

\affiliation{
$^1$ Department of Mathematics, The Ohio State University, Columbus, OH 43210\\
$^2$Department of Physics, University of Connecticut, Storrs, CT 06269}

\begin{abstract}
We show how to convert divergent series, which typically occur in many applications in physics, into rapidly convergent inverse factorial series. This can be interpreted physically as a novel resummation of perturbative series. Being convergent, these new series allow rigorous extrapolation from an asymptotic region with a large parameter, to the opposite region where the parameter is small. We illustrate the method with various physical examples, and discuss how these convergent series relate to standard methods such as Borel summation, and also how they incorporate the physical Stokes phenomenon. We comment on the relation of these results to Dyson's physical argument for the divergence of perturbation theory. This approach also leads naturally to a wide class of relations between bosonic and fermionic partition functions, and Klein-Gordon and Dirac determinants.

\end{abstract}



\maketitle

Physics often requires approximation in the form of an expansion in which a physical parameter becomes large or small compared to other scales. This is a generic feature of perturbation theory, and much physical information is encoded in the relation between the characteristic divergence of perturbation theory and non-perturbative physics \cite{LeGuillou:1990nq,carlbook}. It is less well-known that divergent series can readily be converted to convergent expressions, such as continued fractions  or inverse factorial series \cite{watson,norlund,wall}. For example, a series in inverse powers of a large parameter $x$ may be re-written as a series in inverse Pochhammer symbols, $(x)_{m+1}\equiv \Gamma(x+m+1)/\Gamma(x)=x(x+1)\dots(x+m)$:
\begin{eqnarray}
f(x)=\sum_{n=0}^\infty \frac{c_n}{x^{n+1}} 
=\sum_{m=0}^\infty \sum_{l=0}^m (-1)^{l+m} \frac{S^{(1)}(m,l) \, c_l}{(x)_{m+1}}
\label{eq:inverse}
\end{eqnarray}
where $S^{(1)}(m,l)$ are Stirling numbers of the first kind \cite{nist}. This can be viewed as a resummation of the series expansion, and often this resummed series converges, even if the original series diverges \cite{watson}. Unfortunately, convergence is typically slow, and such series are not accurate near a Stokes line, so they have found somewhat limited use in applications \cite{weniger}. Here we describe physical applications of a new form of inverse factorial series \cite{costins} which is both rapidly convergent and valid near a Stokes line.

To illustrate the basic idea, consider the digamma function, $\psi(1+x)=\frac{d}{dx} \ln \Gamma(1+x)$, which occurs frequently in physics. This has a well-known asymptotic expansion for large argument \cite{nist}, which can be converted to a slowly convergent inverse factorial series using (\ref{eq:inverse}). Instead, consider the integral representation
\begin{eqnarray}
\psi(1+x)=\ln x+\int_0^\infty dp\, e^{-x\, p} \left(\frac{1}{p}-\frac{1}{e^p-1}\right)
\label{eq:psi}
\end{eqnarray}
The remarkable "dyadic identity" (for all $p\in \mathbb C/ \{0\}$)
\begin{eqnarray}
\frac{1}{p}=\frac{1}{(e^{p}-1)}+\sum_{k=1}^\infty \frac{1}{2^k}\frac{1}{(e^{p/2^k}+1)}
\label{eq:id1}
\end{eqnarray}
and successive integrations by parts \footnote{An efficient method is the {\it Horn expansion} \cite{norlund}: given a Borel integral $f(x)=\int_0^\infty dp\, e^{-x\, p} F(p)$, change variables to $s=e^{-p}$, so that $f(x)=\int_0^1 ds\, s^{x-1} \varphi(s)$, where $\varphi(s)\equiv F(-\ln s)$. Then integrating by parts successively one finds the inverse factorial series expansion: $f(x)=\sum_{m=0}^\infty (-1)^m \varphi^{(m)}(1)/(x)_{m+1}$, with coefficients expressed in terms of derivatives of $\varphi(s)$ evaluated at $s=1$. Even when these coefficients cannot be found in closed form it is a simple matter to construct a table of coefficients.} leads to a ``dyadic'' inverse factorial series expansion
\begin{eqnarray}
\psi(1+x)=\ln x+\sum_{k=1}^\infty \sum_{m=0}^\infty \frac{m!}{2^{m+1} (2^k x+1)_{m+1}}
\label{eq:psi2}
\end{eqnarray}
in which the large parameter $x$ is rescaled by factors of $2^k$.
This expansion has the dual advantage of being rapidly convergent, and also  having the correct Stokes jump as the phase of $x$ is rotated. In fact, it gives a global representation valid everywhere in the  cut plane. Fig. 1 shows the convergence even at extremely small values of $x$, contrasted with the truncated asymptotic expansions. 
\begin{figure}[h!]
\centering
\includegraphics[scale=0.84]{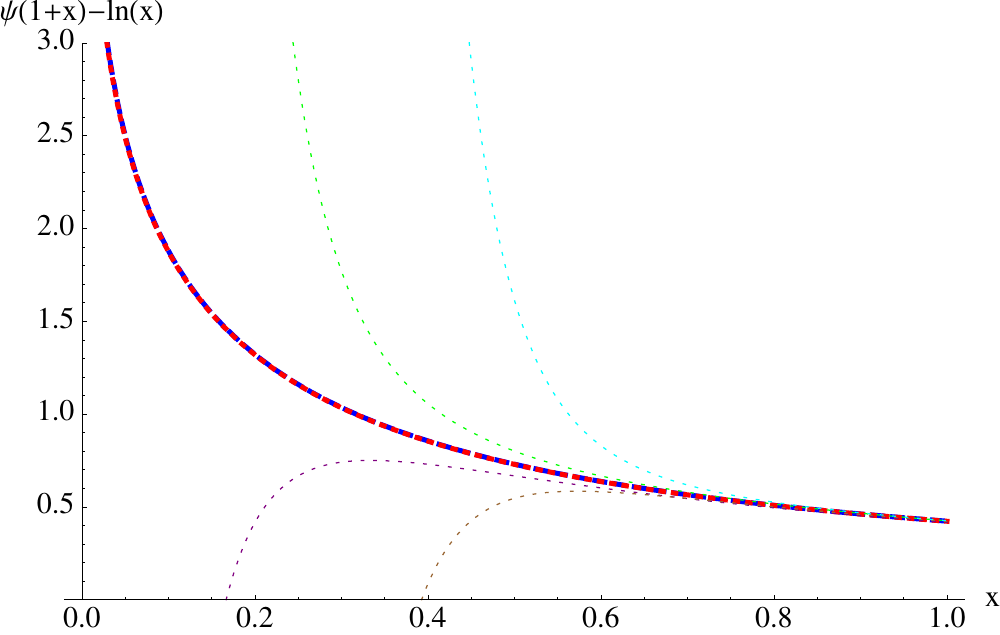} 
\caption{Exact $(\psi(1+x)-\ln x)$ and large $x$ dyadic expansion (\ref{eq:psi2}) (dashed red and blue curves), contrasted with the first 4 partial sums of the standard asymptotic expansion (dotted curves). Even at very small $x$, the dyadic expansion  (\ref{eq:psi2}) is very accurate, while the asymptotic expansions show the typical behavior of diverging rapidly at small $x$.}
\label{fig:psifig}
\end{figure}

This example generalizes to the physical example of the one-loop Euler-Heisenberg (EH) effective Lagrangian \cite{eh,schwinger}, which encodes vacuum polarization effects in quantum electrodynamics (QED). (In fact the EH effective Lagrangian can be expressed in terms of the log of the Barnes gamma function, which involves an integral of the log gamma function  \cite{dunne}, so these examples are closely related.) In the weak field limit the exact (Borel) integral representation of the EH effective Lagrangian is expanded as a divergent series: 
\begin{eqnarray} 
\frac{{\mathcal L}(x)}{m^4}&=& -\frac{1}{8\pi^2 x^2} \int_0^\infty \frac{dp}{p^2} 
e^{-x p}\left(\coth p-\frac{1}{p}-\frac{p}{3}\right) 
\label{eq:eh1}\\
&\sim & -\frac{2}{\pi^2 x^4} \sum_{n=0}^\infty \frac{2^{2n}\, {\mathcal B}_{2n+4}}{(2n+2)(2n+3)(2n+4)} \frac{1}{x^{2n}}
\label{eq:eh2}
\end{eqnarray} 
Here $x\equiv m^2/(e B)$, with $m$ and $e$ the electron mass and charge, and  $B$ the external magnetic field strength.
But $\left(\coth p-\frac{1}{p}-\frac{p}{3}\right)=\sum_{k=1}^\infty 2^{-k} \left(\tanh(p/2^k)-p/2^k\right)$, so
\begin{eqnarray} 
\frac{{\mathcal L}(x)}{m^4} = -\frac{1}{8\pi^2 x^2} \sum_{k=1}^\infty \frac{1}{4^k} \int_0^\infty \frac{dp}{p^2} e^{-2^k x p}\left(\tanh p-p\right) 
\label{eq:eh3}
\end{eqnarray}
\begin{figure}[h!]
\centering
\includegraphics[scale=.7]{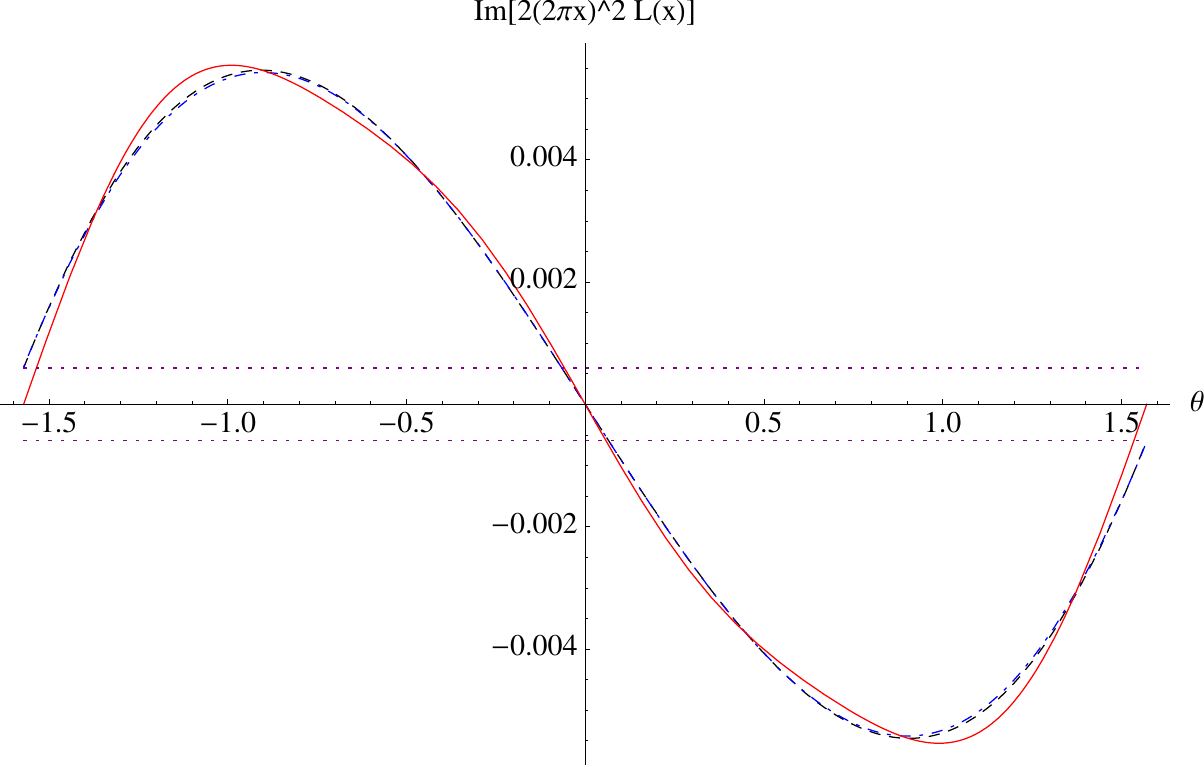}
\caption{Imaginary part of $-8\pi^2 x^2 {\mathcal L}(x)$, with $x=2\, e^{i\theta}$, exact closed form [black dashed curve], the dyadic expression [blue dashed curve] (with just 3 terms of the dyadic $k$ sum, and 6 of the other sum), and from the optimally truncated asymptotic series [red solid curve], as the phase of the parameter $x$ varies in $[-\frac{\pi}{2}, \frac{\pi}{2}]$, changing the magnetic field to an electric field. The Stokes jump (\ref{eq:schwinger}) associated with vacuum pair production is indicated by the horizontal dotted lines. Notice that the dyadic expansion agrees with the closed form expression (in terms of the log of the Barnes Gamma function \cite{dunne}) for all phases, and correctly captures the exponentially small Stokes jump, while the asymptotic series expansion does not.
}
\label{fig:ehfig}
\end{figure}
This generates a new inverse factorial expansion, with the asymptotic parameter $x$ effectively rescaled by $2^k$ for each term in the $k$ sum. 
A convergent dyadic expansion of (\ref{eq:eh3}) is obtained by systematically removing the ``instanton pole'' terms from the integrand, and matching near $p=0$ \footnote{This yields an approximation of the form: $
\frac{\tanh p -p}{p^2}\approx \frac{-8/\pi^2}{1+e^{-2p}}+\frac{4}{\pi^2}+\left(\frac{4}{\pi^2}-\frac{1}{3}\right)p+\dots
$.}. The resulting dyadic weak-field expansion behaves very much like the digamma example plotted in Fig. \ref{fig:psifig}.
The imaginary part is more interesting, and is shown in Fig. \ref{fig:ehfig}. Fig. \ref{fig:ehfig} shows that when the phase of the magnetic field $B$ is rotated, to become an {\it electric} field ($B\to e^{\pm i \pi/2} E$) \footnote{For a constant background field which is either magnetic or electric, the only Lorentz invariant quantity is $(B^2-E^2)$, so $B\to e^{\pm i \pi/2} E$ converts from a background magnetic field to a background electric field.}, the dyadic factorial series encodes the correct imaginary non-perturbative contribution which is associated with the genuine physical process of particle production from the QED vacuum:
\begin{eqnarray}
{\rm Im} \left(\frac{{\mathcal L}\left(\frac{m^2}{e E}\right)}{m^4} \right) =\frac{1}{8\pi^3} \left(\frac{e E}{m^2}\right)^2 {\rm Li}_2\left(e^{-m^2 \pi/(e E)}\right)
\label{eq:schwinger}
\end{eqnarray}

At first sight, this result appears to be in conflict with Dyson's physical argument \cite{Dyson:1952tj} that QED perturbation theory (as an expansion in the fine structure constant $\alpha$) should be divergent, because if it were convergent it would not be able to describe the expected physical instability as the phase of $\alpha=\frac{e^2}{\hbar c}$ is changed from $0$ to $\pm \pi$. 
In mathematical terms, Dyson's argument is that a convergent series in $\alpha$ cannot describe a non-perturbative Stokes jump. The resolution of this apparent contradiction is that Dyson's argument refers to a series expansion in powers of $\alpha=1/x$, whereas the above examples show clearly that a dyadic factorial series correctly describes this non-perturbative effect, even though the dyadic factorial series is convergent. The key fact is that the inverse Pochhammer symbols $1/(x)_n$ of the dyadic factorial expansion encode the proper analytic continuation properties, while the powers $1/x^{n}$ of the divergent expansion do not, even when  truncated at optimal order.

In these two examples, the integral representations (\ref{eq:psi}) and (\ref{eq:eh1}) are in Borel form, with Borel transforms having an infinite sequence of equally spaced poles on the imaginary axis. Physically, the sum over poles in (\ref{eq:eh1}) becomes an instanton sum when the phase of $x$ rotates by $\pm \pi/2$. But note that the poles in (\ref{eq:eh1}) are at integer multiples of $i\pi$, while those of (\ref{eq:eh3}) are at half-odd-integer multiples of $i \pi$. This is because before rescaling $p$ by $2^k$ we have used the fact that every integer can be expressed as a half-odd-integer times a power $2^k$, exactly once \footnote{This interpretation also leads to generalizations from ``dyadic''  to ``n-adic'' inverse factorial series expansions, with even faster convergence, if desired.}. Thus the dyadic expansion, which is a resummation of standard perturbation theory, also produces a nontrivial rearrangement of the non-perturbative instanton sum.

More general physical problems (e.g., for special functions such as Bessel functions or Painlev\'e functions [see below], or in quantum mechanics, quantum field theory and matrix models \cite{LeGuillou:1990nq,carlbook,costinbook,marino}) involve Borel transforms with cuts in the Borel plane. For these we use a generalized dyadic identity in terms of polylog functions (written here for $s<0$ and $p>0$ \footnote{The analytic continuation properties are given by properties of the polylog functions. An alternative representation is 
$\frac{1}{p^s}=\zeta\left(s, p\right)-\sum_{k=1}^\infty \frac{1}{2^{k s}}\zeta\left(s, \frac{1}{2}+\frac{p}{2^k}\right)$.}):
\begin{eqnarray}
p^{s}=\frac{1}{\Gamma(-s)}\left[{\rm Li}_{s+1}(e^{-p})-
\sum_{k=1}^\infty 2^{k s}{\rm Li}_{s+1}\left(-e^{-p/2^k}\right)\right]
\label{eq:id2}
\end{eqnarray}
Note that  (\ref{eq:id1}) is obtained for $s=-1$.
Mathematically, the dyadic identities (\ref{eq:id1}, \ref{eq:id2}) have a natural interpretation in terms of Umbral Calculus \cite{rota}, as geometric Riemann sum approximations to the ``proper-time'' integral:
\begin{eqnarray}
p^s=\frac{1}{\Gamma(-s)}\int_0^\infty \frac{dt}{t^{1+s}} e^{-p t}
\label{eq:umbral}
\end{eqnarray}

\begin{figure}[h!]
\centering
\includegraphics[scale=.85]{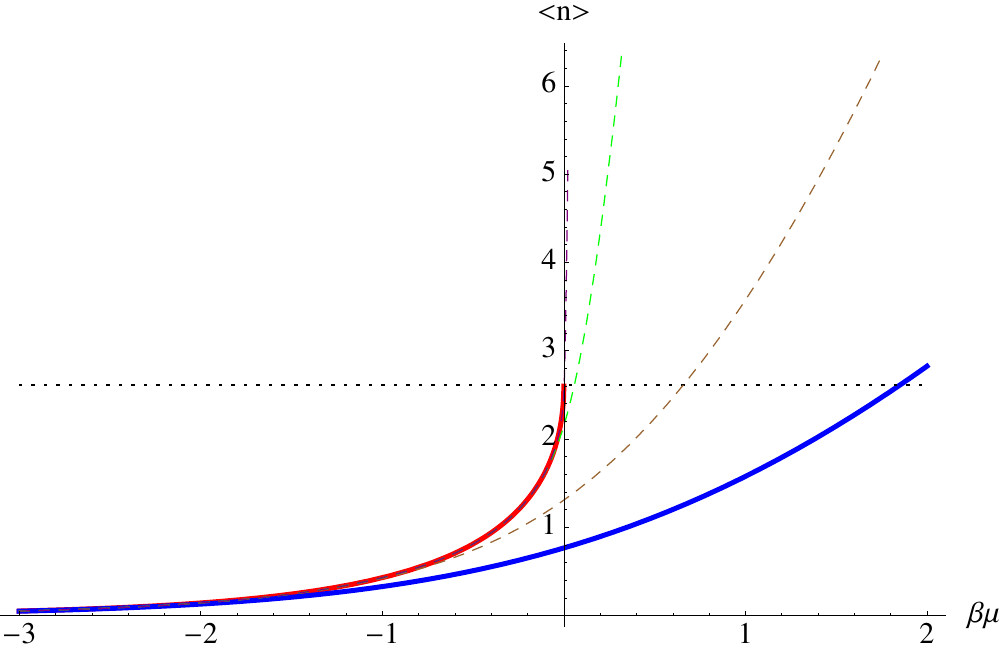}
\caption{Plots of the number density in non-relativistic ideal gases in 3 spatial dimensions. Red curve: bosons; Blue curve: fermions. The horizontal line is the bosonic limiting value ${\rm Li}_{3/2}(1)= \zeta(3/2)$.  The expansion (\ref{eq:nrbf}) is shown for two terms (brown dashed curve), five terms (green dashed curve), and ten terms (purple dashed curve). Note that in the limit the expansion (\ref{eq:nrbf})  correctly describes only $\mu<0$, as is appropriate for bosons, even though it is a sum of fermionic densities.
}
\label{fig:bf}
\end{figure}

The identity (\ref{eq:id1}) can also be interpreted as expressing the Bose-Einstein distribution as a classical term minus a series of Fermi-Dirac distributions with rescaled  temperature. This has numerous thermodynamic consequences. For e.g., thermodynamic quantities for {\it non-relativistic} Bose (+) or Fermi (-) ideal gases are expressed in terms of polylog functions  \cite{pathria,dingle} (here $\beta$ is the inverse temperature, $\mu$ the chemical potential, and $n$ is related to the spatial dimension and the particular physical quantity):
\begin{eqnarray}
f_n^{\pm}(\beta,\mu)=\frac{\beta^{n+1}}{n!}\int _0^\infty \frac{\varepsilon^n \, d\varepsilon}{e^{\beta (\varepsilon-\mu)}\mp 1}
= \pm {\rm Li}_{n+1}\left(\pm e^{\beta \mu}\right)
\label{eq:bf}
\end{eqnarray}
The  number densities in 3 dimensions are: $\lambda_\beta^3 \langle n\rangle_{\rm Bose} = {\rm Li}_{3/2}(e^{\beta \mu})$, and $\lambda_\beta^3 \langle n\rangle_{\rm Fermi} = -{\rm Li}_{3/2}(-e^{\beta \mu})$. Iterating the polylog duplication relation, ${\rm Li}_n(q)+{\rm Li}_n(-q)=2^{1-n}{\rm Li}_n(q^2)$, we learn that the bosonic and fermionic densities are related as (see Fig. \ref{fig:bf}):
\begin{eqnarray}
 \langle n\rangle_{\rm Bose}(\beta \mu) = \sum_{k=0}^\infty \frac{1}{2^{k/2}} \langle n\rangle_{\rm Fermi}\left(2^k \beta\mu\right)
 \label{eq:nrbf}
\end{eqnarray}
Similarly, for {\it relativistic} ideal gases, thermodynamic quantities are expressed in terms of the functions
\begin{eqnarray}
\hskip -6pt g_l^{\pm}(\beta, m, \mu )\hskip -6pt &\equiv& \hskip -6pt \frac{\beta^{l}}{\Gamma(l)} \int_0^\infty \hskip -4pt 
\left[\frac{dq\, q^{l-1}}{e^{\beta(\sqrt{q^2+m^2}-\mu) }\mp 1}-(\mu\to -\mu)\right] 
\nonumber\\
\hskip -6pt h_l^{\pm}(\beta, m, \mu ) \hskip -6pt &\equiv& \hskip -6pt
\frac{\beta^{l-1}}{\Gamma(l)}\int_0^\infty \hskip -4pt
\left[\frac{dq\, q^{l-1}/\sqrt{q^2+m^2}}{e^{\beta(\sqrt{q^2+m^2}-\mu) }\mp 1} +(\mu\to  -\mu)\right] \nonumber
\\
\end{eqnarray} 
 The dyadic identity (\ref{eq:id1}) implies, for example for $l=1$:
\begin{eqnarray}
\hskip -6pt g_1^{+}(\beta, m, \mu ) \hskip -5pt &\equiv& \hskip -5pt \frac{\pi \mu}{\sqrt{m^2-\mu^2}}-\sum_{k=1}^\infty g_1^- \left(\frac{\beta}{2^k}, m, \mu\right)
\nonumber\\
\hskip -6pt h_1^{+}(\beta, m, \mu ) \hskip -5pt &\equiv& \hskip -5pt \frac{\pi/\beta}{\sqrt{m^2-\mu^2}}-\sum_{k=1}^\infty \frac{1}{2^k} h_1^- \left(\frac{\beta}{2^k}, m, \mu\right)
\end{eqnarray} 
(Higher values of the index $l$ can be reached by taking derivatives with respect to $m$ and $\mu$.)
These expressions naturally isolate the non-analytic term $\sqrt{m^2-\mu^2}$ for bosons, whose physical importance for Bose-Einstein condensation has been emphasized in \cite{haber}. These Bose-Fermi relations can also be understood at the level of the partition function using the novel logarithmic identity (obtained from (\ref{eq:id2}) by differentiating wrt $s$ at $s=0$):
\begin{eqnarray}
\ln p = \ln\left(1-e^{-p}\right) -\sum_{k=1}^\infty \ln\left[\frac{1}{2}\left(1+e^{-p/2^k}\right)\right]
\label{eq:ln}
\end{eqnarray}
The dyadic identities (\ref{eq:id1}, \ref{eq:id2}, \ref{eq:ln}) have interesting implications for trans-series expansions \cite{costinbook,Aniceto:2013fka}, wherein a function $F(p)$ is expanded in terms of powers (and possibly iterations) of three basic ``trans-monomial'' objects: $1/p$, $e^{-p}$, and $\ln p$. These are asymptotically independent functions.
However, (\ref{eq:id1}, \ref{eq:id2}, \ref{eq:ln})  show that powers of $p$ and logarithms of $p$ can indeed be expressed in terms of exponentials, but it requires an {\it infinite} number of different exponentials. The logarithmic expression (\ref{eq:ln}) has potential  applications for the computation of entropy, partition functions and effective actions, as an alternative representation of the logarithm compared to conventional proper-time or replica methods.

For example, the effective actions for bosonic or fermionic fields on the hyperbolic manifold $H^2$ are basic building blocks for studying strong-coupling expansions of one-loop corrections for Wilson loop minimal surfaces in $AdS_5\times S^5$ (see, e.g., \cite{forini}). They can be expressed as:
\begin{eqnarray}
\Gamma_{\rm Bose}(m)
\hskip -3pt &=&\hskip -4pt 
\frac{V}{2\pi}\left(\zeta^\prime(-1)+\frac{\ln 2}{12}+ \hskip -4pt \int_0^{m^2+\frac{1}{4}} \hskip -10pt dx\, \psi\left(\frac{1}{2}+\sqrt{x}\right)\right) \nonumber\\
\Gamma_{\rm Dirac}(m) \hskip -3pt &=& \hskip -4pt \frac{V}{\pi}\left(-2\zeta^\prime(-1)-\sqrt{m^2}+  \hskip -4pt \int_0^{m^2} \hskip -10pt 
dx\, \psi\left(1+\sqrt{x}\right)\right) 
\nonumber
\end{eqnarray}
where $V$ is the volume of $H^2$ \cite{forini}.
The dyadic identities, combined with an integral representation of $\psi(x)$, imply another identity
\begin{eqnarray}
\psi(1+x)=2\,\ln 2+\sum_{k=1}^\infty \frac{1}{2^k} \psi\left(\frac{1}{2}+\frac{x}{2^k}\right)
\end{eqnarray}
Therefore, the effective actions are related as:
\begin{eqnarray}
\frac{\pi}{V}\Gamma_{\rm Dirac}(m)&=& -2\zeta^\prime(-1) -\sqrt{m^2}+m^2\,  2\, \ln 2\\
&&\hskip -2cm +\sum_{k=1}^\infty 2^k\left(\frac{2\pi}{V} \Gamma_{\rm Bose}\left(\sqrt{\frac{m^2}{4^k}-\frac{1}{4}}\right) -\zeta^\prime(-1) -\frac{\ln 2}{12}\right)\nonumber
\end{eqnarray}
implying relations  for the strong-coupling expansions.

\begin{figure}[h]
\includegraphics[scale=0.75]{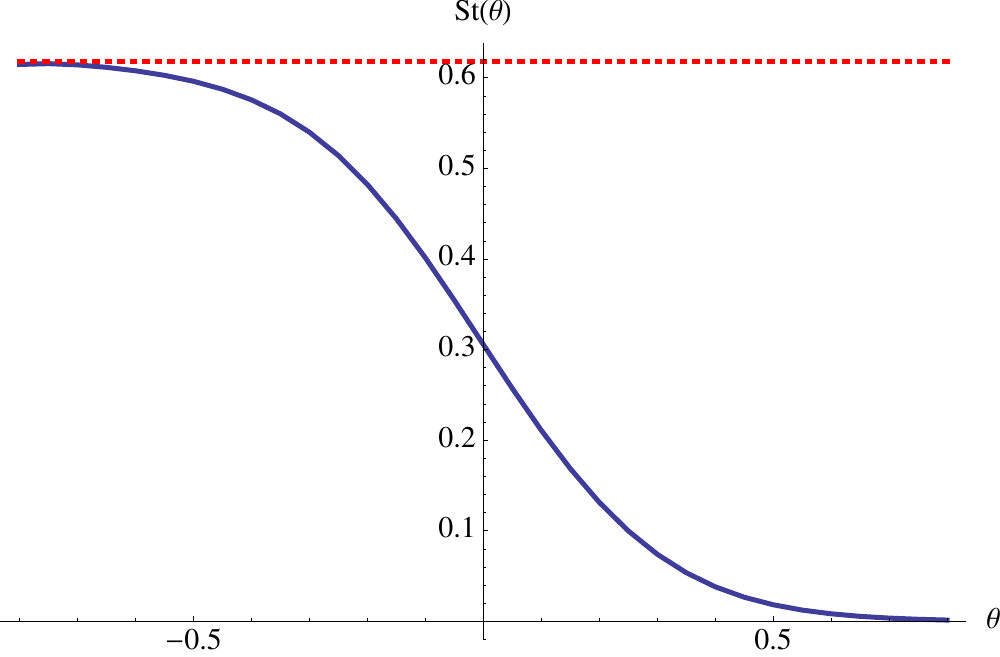}
\caption{The jump across the Stokes line in the convergent dyadic expansion of the solution to the first Painlev\'e equation, as described in the text. The plot shows the imaginary part of the difference between the convergent dyadic expansion and the least-term truncation, ${\rm St}(\theta)=\sqrt{x}e^x(h_{\rm dyadic}(x)-h_{\rm least\,\,term}(x))$, for $x=14\,e^{i\theta}$. Note that as the phase $\theta$ of the variable $x$ is varied across the Stokes line at $\theta=0$, we clearly see the jump corresponding to the magnitude of the Stokes constant: $\sqrt{\frac{6}{5\pi}}=0.618$ (horizontal line). This shows again that the convergent dyadic expansion correctly encodes the Stokes phenomenon.}
\label{fig4}
\end{figure}

So far, we have described examples where an explicit Borel representation is known analytically. However, the convergent dyadic series are also numerically useful when no such explicit Borel transform is known. To illustrate this approach, consider the asymptotic expansion of the solution of the first Painlev\'e equation (${\rm P}_{\rm I}$). The Painlev\'e equations have numerous applications in physics, including random matrix theory, fluid mechanics, statistical physics, matrix models, and 2d quantum gravity \cite{LeGuillou:1990nq,carlbook,clarkson,ising,tracy-widom,forrester,marino,DiFrancesco:1993cyw}. The standard form of ${\rm P}_{\rm I}$, $y^{\prime\prime}(z)=6y^2(z)+z$, can be converted to Boutroux-like form, 
\begin{eqnarray}
h^{\prime\prime}(x)+\frac{1}{x} h^\prime(x)-h(x)-\frac{1}{2} h^2(x)-\frac{392}{625 x^4}=0
\label{eq:boutroux}
\end{eqnarray}
more suitable for asymptotic analysis, by defining $y(z)=i\sqrt{\frac{z}{6}}\left(1-\frac{4}{25x^2}+h(x)\right)$, where $x=e^{i\pi/4} (24 z)^{5/4}/30$ is the natural variable in terms of which the trans-series representation is \cite{costinbook,Garoufalidis:2010ya,costin-huang-tanveer}:
\begin{eqnarray}
h(x)=\sum_{k=0}^\infty C^k \xi^k(x)\, h_{(k)}(x)
\label{eq:trans}
\end{eqnarray}
The sum over $k$ is an ``instanton sum'' with exponentially small ``instanton'' factors  $\xi(x)=e^{-x}/\sqrt{x}$, where $C$ is a trans-series parameter,
and $h_{(k)}(x)$ represents the fluctuations about the $k$-instanton term. It is straightforward to generate many terms in these fluctuation expansions, each of which is a divergent series. We can convert the trans-series expression (\ref{eq:trans})  into a unique global convergent dyadic inverse factorial expansion by the following  procedure \cite{toappear}: (i) From the large-order behavior of the expansion coefficients we learn that the associated Borel transform has two square root branch cuts along the real axis, from $(-\infty, -1]$ and $[+1, +\infty)$; (ii) For each instanton sector, use the conformal map $p=\frac{2z}{z^2+1}$ to map the Borel transform $H_{(k)}(p)$ of $h_{(k)}(x)$ to the unit disc, then re-expand about $z=0$, and re-express in terms of $p$ using the inverse map $z=p/(1+\sqrt{1-p^2})$. This procedure yields an optimal and global representation of the Borel transform in the doubly-cut plane; (iii) Use the Cauchy kernel, and the resurgent properties of the trans-series to express the necessary Laplace transforms as integrals wrapped around the cuts. This step is efficiently done numerically by rotating the cuts to vertical lines emanating from the branch points at integer values of $p$; (iv) Convert the resulting asymptotic expansions to dyadic expansions using the dyadic identity (\ref{eq:id1}). Note that (\ref{eq:id1}) can clearly be generalized by shifting and rescaling, $p\to \alpha +\beta p$, and there is in fact a unique optimal choice of $\beta$ for each instanton series. The output of these algorithmic steps is a unique, global and convergent representation of the function $h(x)$ in the appropriate sector of the physical $x$ plane. As an illustration see Fig. \ref{fig4}, which shows the smoothed Stokes jump \cite{berry,costin-kruskal} in the (convergent) dyadic expansion, as the phase of $x$ is varied across the Stokes line: the jump is equal to the magnitude of the Stokes constant $\sqrt{\frac{6}{5\pi}}=0.618$. We can also extend the analysis, which began in the asymptotic large $x$ regime, to very small values of $x$, and the convergence of the dyadic expansions provides a new simple proof of the Dubrovin conjecture \cite{dubrovin,costin-huang-tanveer} concerning the analyticity of tritronqu\'ee solutions to Painlev\'e-I.
We have illustrated this numerical procedure with Painlev\'e-I, but it is very general, and can be applied systematically to a wide variety of physical problems for which it is possible to generate asymptotic expansions \cite{toappear}.  A novelty of this approach is that it is {\it constructive}, producing a provably convergent and unique expression which still correctly encodes the important non-perturbative physics of the Stokes phenomenon. Furthermore, since these new series are convergent, we may rigorously extrapolate from an asymptotic region with a large parameter to the opposite region of small parameter. This method is therefore ideally adapted to the study of dualities, and general relations between weak-coupling and strong-coupling. The numerical procedure also provides a new way to evaluate Stokes constants with extremely high precision.

\bigskip

This material is based upon work supported by the U.S. Department of Energy, Office of Science, Office of High Energy Physics under Award Number DE-SC0010339 (GD), and by the National Science Foundation under Award Number DMS 1515755 (OC).

\end{document}